\documentclass[aps, prl, nobibnotes, twocolumn, superscriptaddress]{revtex4-2}
\usepackage[T1]{fontenc}
\usepackage{amsmath}
\usepackage{graphicx}
\usepackage{bm}
\usepackage{amssymb}
\usepackage{dcolumn}
\usepackage[colorlinks=true, linkcolor=blue, citecolor=blue,
  urlcolor=blue]{hyperref}

\begin{document}
\title{Piezo-Kerr effect in altermagnets and noncollinear antiferromagnets}

\author{Yuqi Liu}
\thanks{These authors contributed equally to this work.}
\affiliation{Key Lab of advanced optoelectronic quantum architecture and
measurement (MOE), Beijing Key Laboratory of Quantum Matter State Control and
Ultra-Precision Measurement Technology, and School of Physics, Beijing Institute
of Technology, Beijing 100081, China}

\author{Qiu-Shi Huang}
\thanks{These authors contributed equally to this work.}
\affiliation{Key Lab of advanced optoelectronic quantum architecture and
measurement (MOE), Beijing Key Laboratory of Quantum Matter State Control and
Ultra-Precision Measurement Technology, and School of Physics, Beijing Institute
of Technology, Beijing 100081, China}

\author{Xinfeng Chen}
\affiliation{Key Lab of advanced optoelectronic quantum architecture and
measurement (MOE), Beijing Key Laboratory of Quantum Matter State Control and
Ultra-Precision Measurement Technology, and School of Physics, Beijing Institute
of Technology, Beijing 100081, China}

\author{Yilin Han}
\affiliation{Key Lab of advanced optoelectronic quantum architecture and
measurement (MOE), Beijing Key Laboratory of Quantum Matter State Control and
Ultra-Precision Measurement Technology, and School of Physics, Beijing Institute
of Technology, Beijing 100081, China}

\author{Run-Wu Zhang}
\email{zhangrunwu@bit.edu.cn}
\affiliation{Key Lab of advanced optoelectronic quantum architecture and
measurement (MOE), Beijing Key Laboratory of Quantum Matter State Control and
Ultra-Precision Measurement Technology, and School of Physics, Beijing Institute
of Technology, Beijing 100081, China}

\author{Zhi-Ming Yu}
\email{zhiming\_yu@bit.edu.cn}
\affiliation{Key Lab of advanced optoelectronic quantum architecture and
measurement (MOE), Beijing Key Laboratory of Quantum Matter State Control and
Ultra-Precision Measurement Technology, and School of Physics, Beijing Institute
of Technology, Beijing 100081, China}

\author{Yugui Yao}
\affiliation{Key Lab of advanced optoelectronic quantum architecture and
measurement (MOE), Beijing Key Laboratory of Quantum Matter State Control and
Ultra-Precision Measurement Technology, and School of Physics, Beijing Institute
of Technology, Beijing 100081, China}

\begin{abstract}
  Symmetry is usually viewed as a constraint on physical responses, but it can
  also be used to design the fields that control them. Here, we adopt the
  representation-matching principle used to identify conjugate fields of order
  parameters and apply it to response tensors. A linear coupling, the key
  ingredient for sign-switching control, is symmetry-allowed when a component
  (or a combination of components) of the control field and the selected
  response component share the same irreducible representation of the magnetic
  point group (MPG) of the system. Applied to the magneto-optical Kerr effect,
  this principle yields a piezo-Kerr effect, in which specific strain
  combinations activate a Kerr response forbidden in the unstrained state.
  Enumerating all 122 MPGs, we identify 16 groups in which the unstrained Kerr
  response is forbidden while a linear strain-induced contribution is allowed.
  These groups are realized by altermagnets and noncollinear antiferromagnets
  with symmetry-enforced vanishing net magnetization. The active strain channels
  and the associated first-order sign reversals follow directly from symmetry.
  First-principles calculations for bulk MnTe$_{2}$ and CoF$_{2}$, and
  Cr$_{2}$Se$_{2}$O and Cr$_{2}$Te$_{2}$O monolayers confirm these predictions.
  In these materials, the Kerr rotation vanishes without strain, whereas
  0.5--3\% strain produces rotations ranging from tens to hundreds of
  millidegrees, and reversing the matched strain channel reverses the Kerr sign.
  These results establish a general route to large and switchable Kerr responses
  emerging from symmetry-forbidden states in compensated magnets.
\end{abstract}

\maketitle

\emph{\textcolor{blue}{Introduction.--}}The magneto-optical Kerr effect is
central to magneto-optical devices and magnetic information storage, owing to
its noncontact detection of magnetic order through the rotation of
reflected-light polarization \cite{main001, main002, main003, main004, main005}.
Traditionally, studies have focused on ferromagnets, where the net magnetic
moment together with spin-orbit coupling (SOC) can generate a nonzero Kerr
effect \cite{main006, main007, main008, main009}. The Kerr angle can be
controlled by various means, including magnetic fields \cite{main010, main011},
temperature \cite{main010, main012}, current-induced spin torques \cite{main013,
main014}, and gate electric fields \cite{main015, main016, main017}. Switchable
control of the Kerr effect is particularly important for applications
\cite{main018, main019, main020}. However, such sign switching is difficult in
conventional ferromagnets, as an external field mainly modifies the magnitude of
the unperturbed Kerr response rather than reversing its sign.

\begin{figure}[t]
  \includegraphics[width=\columnwidth]{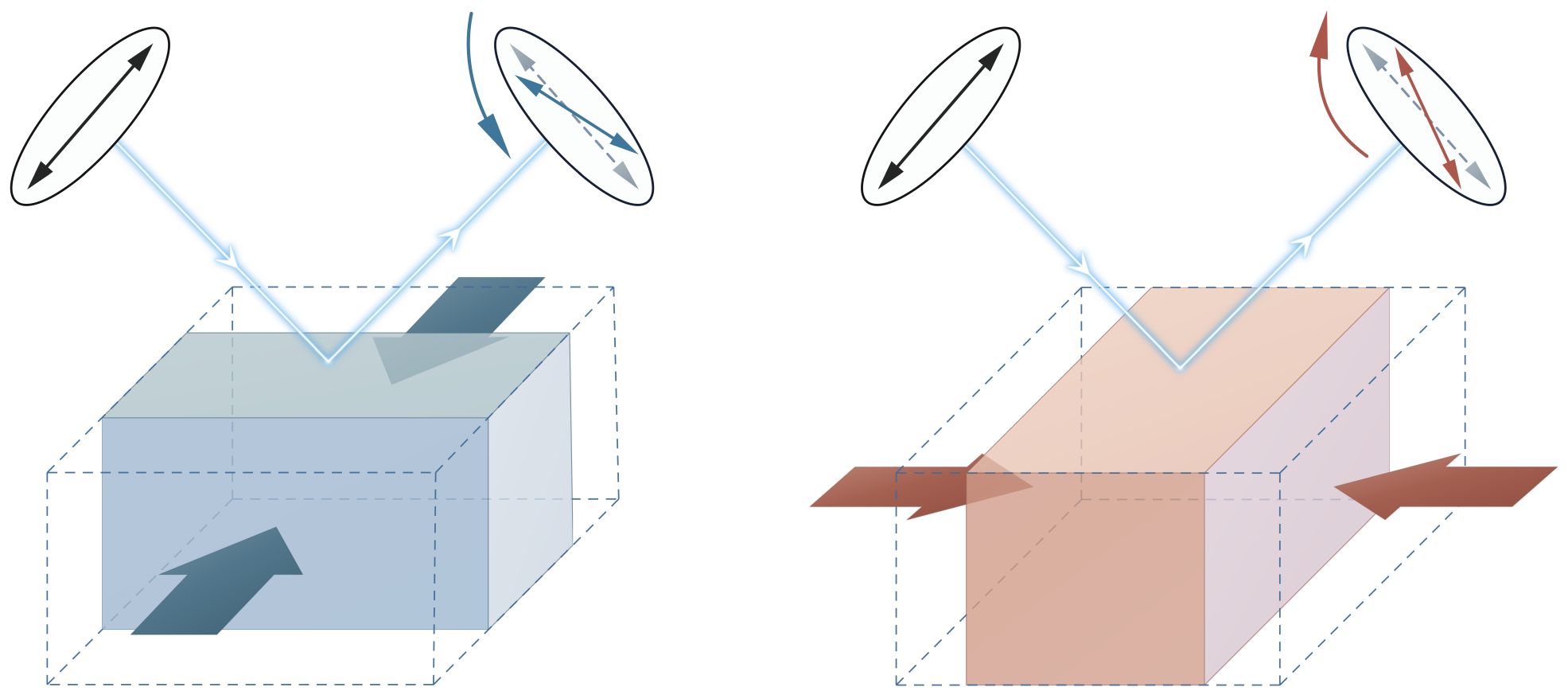}
  \caption{Schematic of the piezo-Kerr effect.
  Specific strain combinations activate a linear Kerr response forbidden in the
  unstrained state. Reversing the matched strain channel reverses the sign of
  the induced Kerr response.
  The matched strains in the schematic are $\varepsilon_{xx}$ and
  $\varepsilon_{yy}$.}
  \label{fig:1}
\end{figure}

Recently, studies of the Kerr effect have been extended to antiferromagnets
(AFMs) \cite{main021, main022, main023, main024} and altermagnets (AMs)
\cite{main025, main026, main027} with vanishing net magnetic moments. When this
vanishing net magnetization is enforced by symmetry, the Kerr effect must also
vanish, because the Kerr response and the net magnetization obey the same
symmetry constraints \cite{main018, main028, main029, main030, main031,
main032}. These compensated materials, by contrast, are ideal platforms for
switchable control of the Kerr effect \cite{main033, main034}. Because the
intrinsic Kerr response vanishes, an external field can both activate the
response and determine its sign, provided a linear coupling between the external
field and the Kerr response is symmetry-allowed.

\renewcommand{\arraystretch}{1.25}
\begin{table*}
  \caption{MPGs allowing the piezo-Kerr effect. Here, $\theta_i$ denotes the
  Kerr response, $\varepsilon_{jk}$ is a component of the strain tensor, and
  $\alpha, \beta, \gamma, \delta$ are material-dependent complex coefficients
  of the piezo-Kerr tensor $\Lambda$.}
  \label{tab:hall_xyz}
  \begin{ruledtabular}
    \begin{tabular}{lcccc}
      MPG &
      \shortstack{incident light $\parallel
      x$\\$\theta_x\propto\sigma_{yz}^{\mathrm A}$} &
      \shortstack{incident light $\parallel
      y$\\$\theta_y\propto\sigma_{zx}^{\mathrm A}$} &
      \shortstack{incident light $\parallel
      z$\\$\theta_z\propto\sigma_{xy}^{\mathrm A}$} &
      \shortstack{Independent\\coefficients} \\
      \colrule
      $222$, $mm2$, $mmm$ & $\alpha\varepsilon_{yz}$ & $\beta\varepsilon_{xz}$ &
      $\gamma\varepsilon_{xy}$ & 3 \\
      \colrule
      $4'$, $\bar4'$, $4'/m$ & $\alpha\varepsilon_{xz}+\beta\varepsilon_{yz}$ &
      $\beta\varepsilon_{xz}-\alpha\varepsilon_{yz}$ &
      $\gamma(\varepsilon_{xx}-\varepsilon_{yy})+\delta\varepsilon_{xy}$ & 4 \\
      \colrule
      $4'22'$, $\bar4'2m'$ & $\alpha\varepsilon_{yz}$ & $\alpha\varepsilon_{xz}$
      & $\beta\varepsilon_{xy}$ & 2 \\
      \colrule
      $4'm'm$, $\bar4'2'm$, $4'/mm'm$ & $\alpha\varepsilon_{xz}$ &
      $-\alpha\varepsilon_{yz}$ & $\beta(\varepsilon_{xx}-\varepsilon_{yy})$ & 2
      \\
      \colrule
      $23$, $m\bar3$, $4'32'$, $\bar4'3m'$, $m\bar3m'$ &
      $\alpha\varepsilon_{yz}$ & $\alpha\varepsilon_{xz}$ &
      $\alpha\varepsilon_{xy}$ & 1 \\
    \end{tabular}
  \end{ruledtabular}
\end{table*}

Mechanical strain is a widely used and well-established means of tuning material
properties and order parameters, such as the piezoelectric \cite{main035,
main036}, piezomagnetic \cite{main037, main038}, piezoaltermagnetic
\cite{main039, main040}, and piezochiral effects \cite{main041}. Strain has also
recently been exploited to control the Kerr effect \cite{main042, main043,
main044}. For AFMs with $\mathcal{PT}$ (spatial inversion combined with time
reversal $\mathcal T$) or $\{\mathcal{T}|\tau\}$ ($\mathcal T$ combined with a
fractional lattice translation) symmetry, however, a homogeneous strain cannot
activate a Kerr response, because it always preserves these symmetries. In
contrast, AMs \cite{main045, main046, main047, main048, main049, main050,
main051, main052, main053, main054, main055, main056, main057, main058, main059,
main060, main061, main062} are free of this restriction. Yet the symmetry
conditions and the explicit form of the linear coupling between strain and the
Kerr response, which determine the switchable control of the Kerr effect, have
not been established. Moreover, the strain-Kerr coupling in noncollinear AFMs is
likewise unexplored.

Here, we address this problem by developing a symmetry framework for the
piezo-Kerr effect, a Kerr response that is symmetry-forbidden in the unstrained
state but activated by strain. A linear strain-Kerr coupling is symmetry-allowed
when a strain component (or a combination of components) shares the same
irreducible representation (IRR) of the magnetic point group (MPG) as the Kerr
response. By this representation-matching criterion, we enumerate all 122 MPGs
and identify 16 groups in which the unstrained Kerr response is forbidden while
a linear strain-induced contribution remains allowed. All 16 groups are realized
by AMs and noncollinear AFMs with symmetry-enforced vanishing net magnetization.
First-principles calculations for bulk MnTe$_{2}$ and CoF$_{2}$, and
Cr$_{2}$Se$_{2}$O and Cr$_{2}$Te$_{2}$O monolayers confirm these predictions. In
these materials, the Kerr rotation vanishes without strain, whereas moderate
strain produces rotations of tens to hundreds of millidegrees, and reversing the
matched strain reverses the Kerr sign (Fig.~\ref{fig:1}). These results
establish mechanical strain as a general, reversible means for controlling Kerr
responses in compensated magnets, with direct implications for switchable
magneto-optical devices.

\emph{\textcolor{blue}{Symmetry of the piezo-Kerr effect.--}}Let $\theta_i$
denote the polar Kerr rotation for light propagating along the Cartesian
direction $i$ ($i=x,y,z$). Since $\theta_i$ is proportional to the antisymmetric
part of the optical conductivity, $\theta_i\propto\sigma^{\mathrm A}_{jk}$ with
$(i,j,k)$ a cyclic permutation of $(x,y,z)$, $\bm\theta$ is a $\mathcal T$-odd
axial vector. The piezo-Kerr effect here is defined as the linear coupling
between a homogeneous strain $\varepsilon_{jk}$ and $\theta_i$,
\begin{equation}
  \theta_i=\sum_{jk}\Lambda_{i,jk}\,\varepsilon_{jk},
\end{equation}
where $\Lambda$ is the piezo-Kerr tensor. Because strain $\varepsilon$ is
$\mathcal T$-even, the tensor $\Lambda$ is $\mathcal T$-odd.

The Kerr rotation is conventionally controlled by a magnetic field, because the
magnetic field is $\mathcal T$-odd. Strain, in contrast, is $\mathcal T$-even
and is generally not expected to couple directly to the Kerr effect. This is
true for nonmagnetic systems, where $\mathcal T$ reverses $\theta_i$ but leaves
$\varepsilon$ unchanged. When $\mathcal T$ is broken, however, a
symmetry-matched field other than the magnetic field can emerge for $\bm\theta$.
For a system with a given MPG $\mathcal M$, strain can serve as such a
symmetry-matched field precisely when it shares with the Kerr response the same
IRR of $\mathcal M$. This can be realized in the systems without $\mathcal T$.

As an example, in a system whose only symmetry is $\mathcal C_{4z}\mathcal T$,
$\mathcal C_{4z}$ leaves $\theta_z$ unchanged but reverses
$\varepsilon_{xx}-\varepsilon_{yy}$, whereas $\mathcal T$ reverses $\theta_z$
but leaves $\varepsilon_{xx}-\varepsilon_{yy}$ unchanged. The combined operation
therefore reverses both, so $\theta_z$ and $\varepsilon_{xx}-\varepsilon_{yy}$
share the same representation, making the linear coupling
$\theta_z\propto\varepsilon_{xx}-\varepsilon_{yy}$ symmetry-allowed.

Formally, this representation-matching condition is expressed by the constraint
\begin{equation}
  \Lambda=D_\theta(g)\Lambda D_\varepsilon^{-1}(g),
\end{equation}
imposed by each symmetry operation $g$ of $\mathcal M$. Here $D_\theta(g)$ and
$D_\varepsilon(g)$ are the transformation (representation) matrices of $\theta$
and $\varepsilon$ under $g$, respectively. This constraint completely determines
the form of $\Lambda$ \cite{main030, main063}. We further illustrate this with
the point groups $D_{4h}$ and $4'/mm'm$. Under $D_{4h}$, $\theta_z$ transforms
as the IRR $A_{2g}$ and $\varepsilon_{xx}-\varepsilon_{yy}$ as IRR $B_{1g}$.
Thus, their coupling is forbidden.
In contrast, for $4'/mm'm$, both $\theta_z$ and
$\varepsilon_{xx}-\varepsilon_{yy}$ transform as the same IRR $B_{1g}$, allowing
the linear coupling between the Kerr response and strain. Enumerating all 122
MPGs in this way, we identify 16 groups in which the unstrained Kerr response is
forbidden, while a linear strain-induced contribution remains allowed, as listed
in Table~\ref{tab:hall_xyz}.

\begin{table*}
  \caption{Representative materials exhibiting the piezo-Kerr effect. The
  materials are grouped into two categories: those with a symmetry-enforced
  vanishing net magnetization (belonging to the 16 MPGs listed in
  Table~\ref{tab:hall_xyz}) and those with an accidentally vanishing net
  magnetization. ``nonzero'' denotes a symmetry-allowed Kerr effect in the
  unstrained state, whereas the other entries give the linear strain-induced
  Kerr effect. Coefficient conventions follow those in Table~\ref{tab:hall_xyz}.
  2D and 3D denote two- and three-dimensional systems.}
  \label{tab:II}
  \begin{ruledtabular}
    \begin{tabular}{@{}llcccl@{}}
      AM or noncollinear AFM & MPG & $\theta_x$ &
      $\theta_y$ & $\theta_z$ & Material \\
      \colrule
      Essential & $222$ & $\alpha\varepsilon_{yz}$ & $\beta\varepsilon_{xz}$ &
      $\gamma\varepsilon_{xy}$ & DyFeO$_{3}$ (3D coplanar AFM) \cite{main064} \\
      & $\bar4'$ & $\alpha\varepsilon_{xz}+\beta\varepsilon_{yz}$ &
      $\beta\varepsilon_{xz}-\alpha\varepsilon_{yz}$ &
      $\gamma(\varepsilon_{xx}-\varepsilon_{yy})+\delta\varepsilon_{xy}$ &
      CsCoF$_{4}$ (3D collinear AFM) \cite{main065} \\
      & $4'/mm'm$ & $\alpha\varepsilon_{xz}$ & $-\alpha\varepsilon_{yz}$ &
      $\beta(\varepsilon_{xx}-\varepsilon_{yy})$ & Cr$_{2}$Se$_{2}$O (2D
      collinear AM) \cite{main066, main067} \\
      & $4'/mm'm$ & $\alpha\varepsilon_{xz}$ & $-\alpha\varepsilon_{yz}$ &
      $\beta(\varepsilon_{xx}-\varepsilon_{yy})$ & Cr$_{2}$Te$_{2}$O (2D
      collinear AM) \cite{main067} \\
      & $4'/mm'm$ & $\alpha\varepsilon_{xz}$ & $-\alpha\varepsilon_{yz}$ &
      $\beta(\varepsilon_{xx}-\varepsilon_{yy})$ & CoF$_{2}$ (3D collinear AM)
      \cite{main068} \\
      & $m\bar3$ & $\alpha\varepsilon_{yz}$ & $\alpha\varepsilon_{xz}$ &
      $\alpha\varepsilon_{xy}$ & MnTe$_{2}$ (3D noncoplanar AFM) \cite{main069,
      main070} \\
      \colrule
      Accidental & $2'/m'$ & nonzero &
      $\alpha\varepsilon_{xy}+\beta\varepsilon_{yz}$ & nonzero &
      $\alpha$-Fe$_{2}$O$_{3}$ (3D coplanar AFM) \cite{main027} \\
      & $2'/m'$ & nonzero & $\alpha\varepsilon_{xy}+\beta\varepsilon_{yz}$ &
      nonzero & AgF$_{2}$ (2D collinear AM) \cite{main071} \\
      & $2'2'2$ & $\alpha\varepsilon_{xz}$ & $\beta\varepsilon_{yz}$ & nonzero &
      Ca$_{2}$CoSi$_{2}$O$_{7}$ (3D coplanar AFM) \cite{main072} \\
      & $2$ & $\alpha\varepsilon_{xy}+\beta\varepsilon_{yz}$ & nonzero &
      $\gamma\varepsilon_{xy}+\delta\varepsilon_{yz}$ & LiFeP$_{2}$O$_{7}$ (3D
      collinear AFM) \cite{main073} \\
      & $2$ & $\alpha\varepsilon_{xy}+\beta\varepsilon_{yz}$ & nonzero &
      $\gamma\varepsilon_{xy}+\delta\varepsilon_{yz}$ & OsNNaSCl$_{5}$ (2D
      collinear AM) \cite{main071} \\
      & $m'm2'$ & $\alpha\varepsilon_{xy}$ & nonzero & $\beta\varepsilon_{yz}$ &
      Ca$_{3}$Mn$_{2}$O$_{7}$ (3D collinear AFM) \cite{main074} \\
    \end{tabular}
  \end{ruledtabular}
\end{table*}

\begingroup\widowpenalty=10000
  \emph{\textcolor{blue}{Material realization.--}}To test the classification
  in real materials, we select four representative magnets whose MPGs belong to
  the 16 groups identified in Table~\ref{tab:hall_xyz}. These include the bulk
  noncoplanar AFM MnTe$_{2}$, the bulk collinear AM CoF$_{2}$, and the collinear
  AM monolayers Cr$_{2}$Se$_{2}$O and Cr$_{2}$Te$_{2}$O, covering two- and
  three-dimensional systems with collinear and noncollinear magnetic orders.
  Table~\ref{tab:II} lists the representative magnets together with their
  piezo-Kerr couplings. All of them have a vanishing net magnetization, but only
  in the 16 MPGs is that enforced by symmetry. In the remaining materials, the
  vanishing net magnetization is accidental. As a consequence, their unstrained
  Kerr response vanishes only along specific directions rather than in
  \mbox{every direction}.
\par\endgroup

\begin{figure}
  \includegraphics[width=\columnwidth]{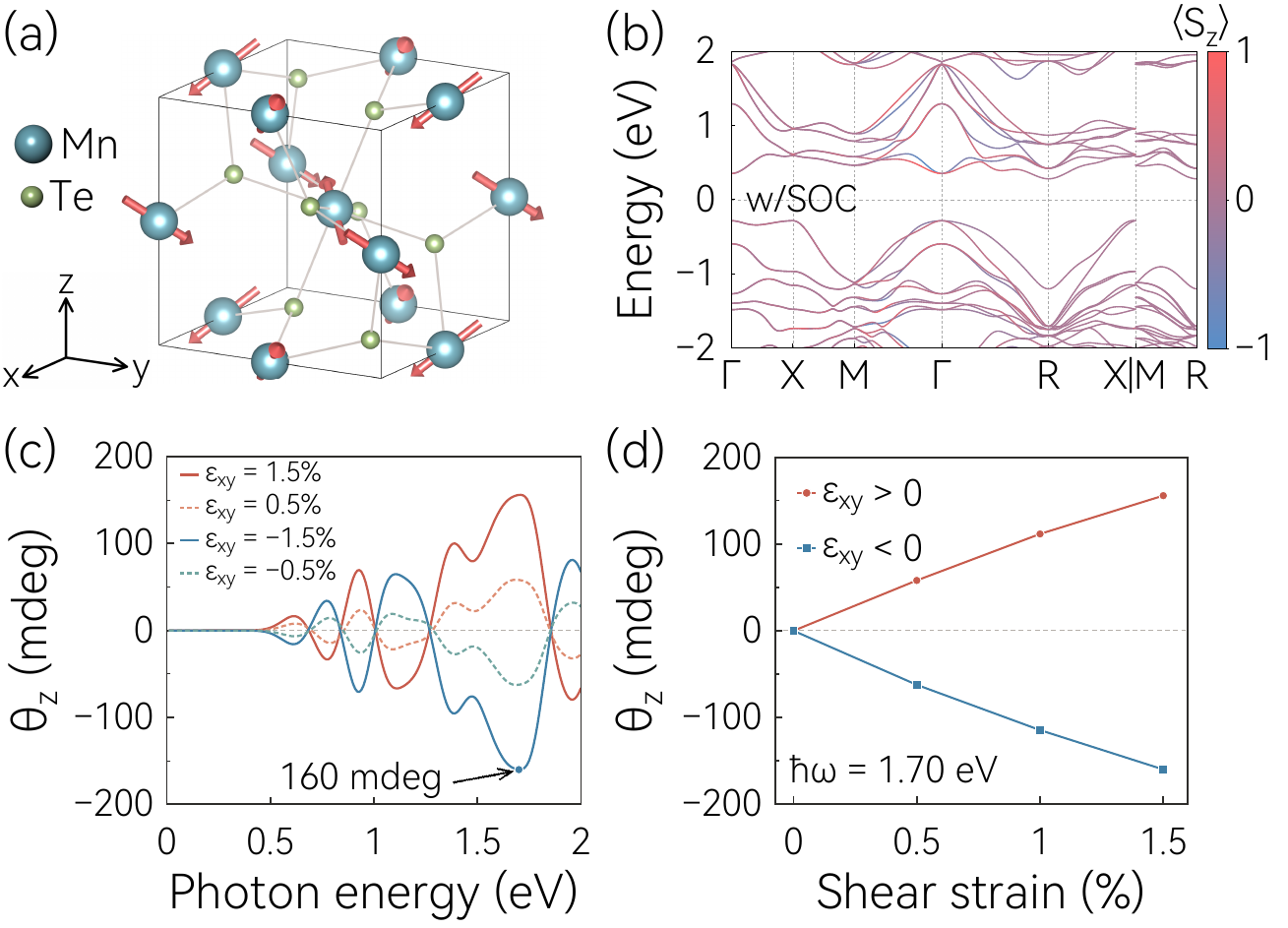}
  \caption{(a) Crystal structure and magnetic configuration of bulk MnTe$_2$.
  Arrows denote magnetic moments. (b) Band structure of MnTe$_2$ with SOC and
  without strain. (c) Kerr rotation $\theta_z$ under different shear strains
  $\varepsilon_{xy}$. (d) Strain dependence of $\theta_z$ at a photon energy of
  $\hbar\omega=1.70$ eV.}
  \label{fig:2}
\end{figure}

Bulk MnTe$_2$ crystallizes in the cubic pyrite structure, with the Mn moments
forming a noncoplanar arrangement along the body diagonals [Fig.~\ref{fig:2}(a)]
\cite{main069, main070, main075}. Its N\'eel temperature is approximately 87 K
\cite{main069}. Its MPG is $m\bar3$. According to Table~\ref{tab:hall_xyz}, the
Kerr rotation $\theta_z$ in MnTe$_2$ couples to the shear strain
$\varepsilon_{xy}$. For unstrained MnTe$_2$, the spin splitting observed
experimentally \cite{main069} is evident in the SOC band structure, which also
shows its semiconducting character [Fig.~\ref{fig:2}(b)]. We focus on the Kerr
rotation $\theta_z$ \cite{main007, main076, main077}, which is zero by symmetry
without strain. Under 0.5\% $xy$ shear strain, $\theta_z$ becomes finite and
reverses sign when the strain is reversed [Fig.~\ref{fig:2}(c)]. To facilitate
experimental detection, we examine the strain dependence at a fixed photon
energy near a pronounced Kerr-rotation peak for each material. At a fixed
photon energy of $\hbar\omega=1.70$ eV, $|\theta_z|$ grows with increasing
shear-strain magnitude, reaching approximately 160 mdeg at 1.5\% shear strain
[Fig.~\ref{fig:2}(d)].

\begin{figure}
  \includegraphics[width=\columnwidth]{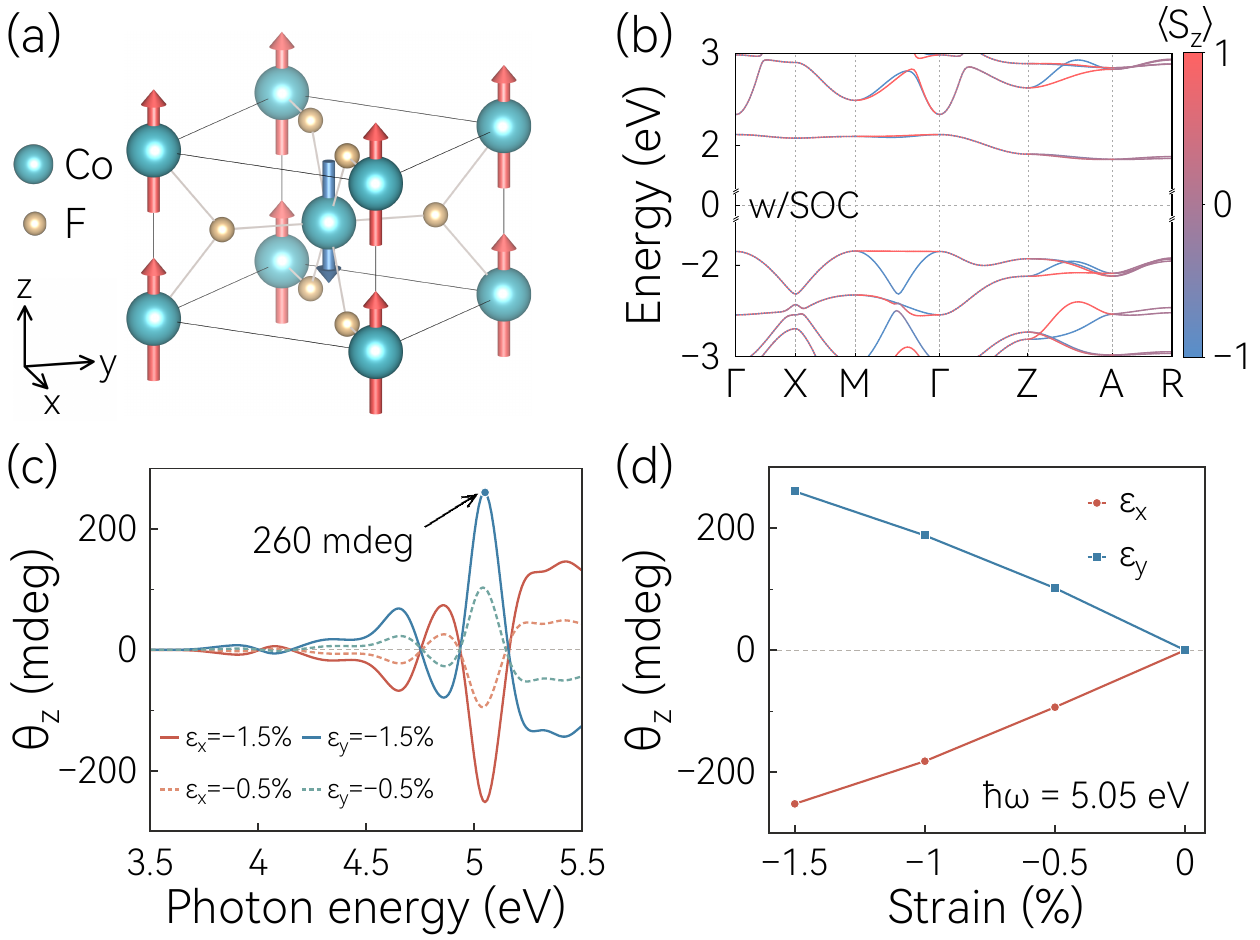}
  \caption{(a) Crystal structure and magnetic configuration of bulk CoF$_2$.
  Arrows denote magnetic moments. (b) Band structure with SOC and without
  strain. (c) Kerr rotation $\theta_z$ under uniaxial compression along $x$ and
  $y$ directions. (d) Strain dependence of $\theta_z$ at a photon energy of
  $\hbar\omega=5.05$ eV.}
  \label{fig:3}
\end{figure}

Bulk CoF$_2$ is a collinear AM with a tetragonal rutile structure, in which the
two Co sublattices carry antiparallel magnetic moments along the $z$ axis
[Fig.~\ref{fig:3}(a)] \cite{main078, main068, main079}. Its MPG is $4'/mm'm$,
one of the 16 groups listed in Table~\ref{tab:hall_xyz}. The experimentally
reported N\'eel temperature of CoF$_2$ is 39 K \cite{main068}.
According to Table~\ref{tab:hall_xyz}, the Kerr rotation $\theta_z$ couples to
the strain $\varepsilon_{xx}-\varepsilon_{yy}$. The calculated unstrained band
structure of CoF$_2$ with SOC ($\bm N\parallel z$, where $\bm N$ is the N\'eel
vector) is shown in Fig.~\ref{fig:3}(b), revealing its insulating character and
momentum-dependent spin splitting. Symmetry requires $\theta_z=0$ in unstrained
CoF$_2$. A uniaxial compressive strain of 0.5\% induces a finite $\theta_z$,
which changes sign when the strain axis is switched between $x$ and $y$
[Fig.~\ref{fig:3}(c)]. With increasing compressive strain, $|\theta_z|$ at a
fixed photon energy of $\hbar\omega=5.05$ eV grows, reaching approximately
260 mdeg at 1.5\% strain [Fig.~\ref{fig:3}(d)].

Monolayer Cr$_{2}$Se$_{2}$O \cite{main066, main067, main080} and
Cr$_{2}$Te$_{2}$O \cite{main067} share the same crystal structure and magnetic
configuration, shown in Fig.~\ref{fig:4}(a) and Fig.~S1(a) of the Supplemental
Material (SM) \cite{main081}\nocite{main082, main083, main084, main085, main086,
main087, main088, main089, main090, main091, main092, main093, main094, main095,
main096, main097}, respectively. They crystallize in the tetragonal space group
$P4/mmm$, with a central Cr$_{2}$O layer sandwiched between two chalcogen
sublayers (Se in Cr$_{2}$Se$_{2}$O, and Te in Cr$_{2}$Te$_{2}$O). The optimized
in-plane lattice constants are 4.02 and 4.10 \AA{} for Cr$_{2}$Se$_{2}$O and
Cr$_{2}$Te$_{2}$O, respectively. First-principles calculations show that the
magnetic ground state of both monolayers is altermagnetic, with the two Cr
sublattices carrying antiparallel magnetic moments normal to the layer. Thus,
both monolayers have the MPG $4'/mm'm$, in which the Kerr rotation $\theta_z$
couples to the normal-strain difference $\varepsilon_{xx}-\varepsilon_{yy}$, as
listed in Table~\ref{tab:hall_xyz}.

\suppressfloats[t]
\begin{figure}[t]
  \includegraphics[width=\columnwidth]{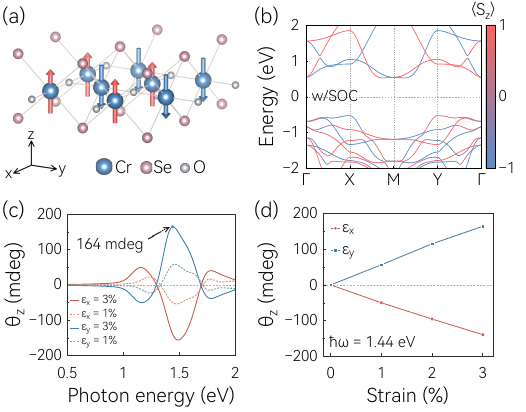}
  \caption{(a) Crystal structure and magnetic configuration of monolayer
  Cr$_{2}$Se$_{2}$O. Arrows denote magnetic moments. (b) Band structure with SOC
  and without strain. (c) Kerr rotation $\theta_z$ under uniaxial tension along
  $x$ and $y$ directions. (d) Strain dependence of $\theta_z$ at a photon energy
  of $\hbar\omega=1.44$~eV.}
  \label{fig:4}
\end{figure}

The calculated unstrained band structure of monolayer Cr$_{2}$Se$_{2}$O with SOC
($\bm N\parallel z$) is shown in Fig.~\ref{fig:4}(b), revealing its
semiconducting character and momentum-dependent spin splitting. In the absence
of strain, $\theta_z$ vanishes by symmetry. Under uniaxial tensile strain,
however, $\theta_z$ becomes finite, and its sign reverses when the strain axis
is switched between the $x$ and $y$ axes [Fig.~\ref{fig:4}(c)]. Over a suitable
range of strain magnitude, $\theta_z$ varies linearly with strain
[Fig.~\ref{fig:4}(d)]. This confirms the linear
$\varepsilon_{xx}-\varepsilon_{yy}$ coupling listed in Table~\ref{tab:hall_xyz}.
Remarkably, $|\theta_z|$ at a fixed photon energy of $\hbar\omega=1.44$ eV
reaches $\sim$50 mdeg at 1\% strain for both $x$- and $y$-axis tension, and
164 mdeg at 3\% strain [Fig.~\ref{fig:4}(d)]. The 164 mdeg value is
significantly larger than the calculated Kerr angles of 17.2 mdeg in
gate-activated bilayer MnPSe$_{3}$
\cite{main015} and 100 mdeg in monolayer Cr$_{2}$O$_{2}$ \cite{main043}. For
comparison, the measured Kerr angles are 0.7--2 mdeg in bilayer and trilayer
Cr$_{2}$Ge$_{2}$Te$_{6}$ \cite{main012} and 103.7 mdeg in bulk
$\alpha$-Fe$_{2}$O$_{3}$ \cite{main027}.
The results of monolayer Cr$_{2}$Te$_{2}$O, which show similar behavior, are
presented in the SM \cite{main081}.
The unstrained band structures without SOC and strained band structures with SOC
of all four materials are presented in Figs.~S2--S5 of the SM \cite{main081}.

\emph{\textcolor{blue}{Discussion.--}}
In this work, we introduced the concept of the piezo-Kerr effect and established
a symmetry framework for it. We identified the 16 MPGs that host the piezo-Kerr
effect. First-principles calculations for MnTe$_{2}$, CoF$_{2}$, and monolayers
Cr$_{2}$Se$_{2}$O and Cr$_{2}$Te$_{2}$O confirmed these predictions, giving Kerr
rotations from tens to hundreds of millidegrees under 0.5--3\% strain, with the
sign flipping when the matched strain channel is reversed. Because the
unstrained background is exactly zero, the activated signal is clean and its
sign is fully set by the strain, making compensated magnets a promising platform
for switchable magneto-optical devices.

Although we have focused on the Kerr effect, the representation-matching
principle is general and applies to any response tensor for which a
symmetry-matched field is sought. It directly identifies which field components
can couple linearly to a chosen response component in a given MPG. The
piezo-Kerr effect is the magneto-optical realization of this principle, in which
a $\mathcal T$-even field (strain) controls a $\mathcal T$-odd response (the
Kerr rotation) through the magnetic order.

The predicted Kerr rotations of tens to hundreds of millidegrees are readily
detectable, well above the microdegree sensitivity of modern Kerr measurements
\cite{main098}. The sign reversal tied to the matched strain channel offers a
natural binary switching mechanism, in which the zero unstrained background lets
the strain determine both the magnitude and the sign of the response. The
distinction between symmetry-enforced and accidental vanishing net magnetization
matters here, because only the former guarantees a strictly zero Kerr background
in every direction, making the switchable response robust against perturbations.

Finally, a systematic search within the 16 MPGs, guided by the symmetry-adapted
strain basis in Table~\ref{tab:hall_xyz}, should reveal further candidate
materials and could \mbox{extend} the same design principle to other
magneto-optical and transport responses.

\addvspace{\baselineskip}
\begin{acknowledgments}
  This work is supported by the National Key R\&D Program of China (Grant No.
  2025YFA1411200), the National Natural Science Foundation of China (Grants No.
  12234003, No. 12474040, No. W2511003, No. 12304188, No. 12321004), and the
  Natural Science Foundation of Beijing (Grant No. 1252029).
\end{acknowledgments}

\addvspace{\baselineskip}
\textit{Data availability---}The data are not publicly available. The data are
available from the authors upon reasonable request.

\setlength{\bibsep}{0pt plus 1pt}
\makeatletter
\renewcommand{\bibsection}{\begingroup\baselineskip14pt\relax\bib@device
  {\hsize}{72pt}\endgroup\addvspace{7pt}}
\makeatother
\bibliographystyle{apsrev4-2}
\bibliography{Manuscript/reference}
\end{document}

\frenchspacing
\title{Supplemental Material for \textquotedblleft Piezo-Kerr effect in
altermagnets and noncollinear antiferromagnets\textquotedblright}
\author{Yuqi Liu}
\thanks{These authors contributed equally to this work.}
\affiliation{Key Lab of advanced optoelectronic quantum architecture and
measurement (MOE), Beijing Key Laboratory of Quantum Matter State Control and
Ultra-Precision Measurement Technology, and School of Physics, Beijing Institute
of Technology, Beijing 100081, China}

\author{Qiu-Shi Huang}
\thanks{These authors contributed equally to this work.}
\affiliation{Key Lab of advanced optoelectronic quantum architecture and
measurement (MOE), Beijing Key Laboratory of Quantum Matter State Control and
Ultra-Precision Measurement Technology, and School of Physics, Beijing Institute
of Technology, Beijing 100081, China}

\author{Xinfeng Chen}
\affiliation{Key Lab of advanced optoelectronic quantum architecture and
measurement (MOE), Beijing Key Laboratory of Quantum Matter State Control and
Ultra-Precision Measurement Technology, and School of Physics, Beijing Institute
of Technology, Beijing 100081, China}

\author{Yilin Han}
\affiliation{Key Lab of advanced optoelectronic quantum architecture and
measurement (MOE), Beijing Key Laboratory of Quantum Matter State Control and
Ultra-Precision Measurement Technology, and School of Physics, Beijing Institute
of Technology, Beijing 100081, China}

\author{Run-Wu Zhang}
\email{zhangrunwu@bit.edu.cn}
\affiliation{Key Lab of advanced optoelectronic quantum architecture and
measurement (MOE), Beijing Key Laboratory of Quantum Matter State Control and
Ultra-Precision Measurement Technology, and School of Physics, Beijing Institute
of Technology, Beijing 100081, China}

\author{Zhi-Ming Yu}
\email{zhiming\_yu@bit.edu.cn}
\affiliation{Key Lab of advanced optoelectronic quantum architecture and
measurement (MOE), Beijing Key Laboratory of Quantum Matter State Control and
Ultra-Precision Measurement Technology, and School of Physics, Beijing Institute
of Technology, Beijing 100081, China}

\author{Yugui Yao}
\affiliation{Key Lab of advanced optoelectronic quantum architecture and
measurement (MOE), Beijing Key Laboratory of Quantum Matter State Control and
Ultra-Precision Measurement Technology, and School of Physics, Beijing Institute
of Technology, Beijing 100081, China}
\maketitle
\section{STRAIN REPRESENTATION AND SYMMETRY CONSTRAINTS}

\subsection{Strain tensor and linear coupling}

The strain components entering Eq. (1) of the main text are expressed in fixed
Cartesian axes. The indices $i,j,k$ each take the values $x,y,z$. The
homogeneous strain tensor is symmetric:
\begin{equation}
  \bm\varepsilon=
  \begin{pmatrix}
    \varepsilon_{xx}&\varepsilon_{xy}&\varepsilon_{xz}\\
    \varepsilon_{yx}&\varepsilon_{yy}&\varepsilon_{yz}\\
    \varepsilon_{zx}&\varepsilon_{zy}&\varepsilon_{zz}
  \end{pmatrix},
  \qquad \varepsilon_{jk}=\varepsilon_{kj}.
  \label{eq:sm_strain_tensor}
\end{equation}
It has six independent components: three normal components
$(\varepsilon_{xx},\varepsilon_{yy},\varepsilon_{zz})$ and three shear
components $(\varepsilon_{yz},\varepsilon_{zx},\varepsilon_{xy})$. The
off-diagonal entries are tensor shear strains, with no factor of 2 included in
their definition.

Let $\theta_i$ denote the polar Kerr rotation for light propagating along the
Cartesian direction $i$ ($i=x,y,z$). Expanding Eq. (1) of the main text and
using Eq.~\eqref{eq:sm_strain_tensor} gives
\begin{equation}
  \begin{aligned}
    \theta_i={}&\Lambda_{i,xx}\varepsilon_{xx}+\Lambda_{i,yy}\varepsilon
    _{yy}+\Lambda_{i,zz}\varepsilon_{zz}\\
    &+(\Lambda_{i,yz}+\Lambda_{i,zy})\varepsilon_{yz}
    +(\Lambda_{i,zx}+\Lambda_{i,xz})\varepsilon_{zx}\\
    &+(\Lambda_{i,xy}+\Lambda_{i,yx})\varepsilon_{xy}.
  \end{aligned}
  \label{eq:sm_coupling_expanded}
\end{equation}
Here, $\Lambda$ is the piezo-Kerr tensor. The two terms associated with each
shear component are combined because $\varepsilon_{jk}=\varepsilon_{kj}$. In the
six-component form used in Eq. (2) of the main text, $\Lambda$ is a $3\times6$
coefficient matrix. Its rows correspond to $\theta_x,\theta_y,\theta_z$, and its
columns follow the order $(xx,yy,zz,yz,zx,xy)$. Each shear-column coefficient is
the corresponding sum of tensor components in
Eq.~\eqref{eq:sm_coupling_expanded}.

\subsection{Symmetry constraints}

For a given magnetic point group (MPG) $\mathcal M$, let $R_g$ be the $3\times3$
orthogonal spatial transformation matrix of each symmetry operation
$g\in\mathcal M$ in Eq. (2) of the main text. The index $t_g$ is 0 when $g$ does
not contain time reversal $\mathcal T$, and 1 when it does. A prime denotes the
quantity after applying $g$, expressed in the same Cartesian axes. The
superscript $\mathrm T$ denotes matrix transpose. The determinant $\det(R_g)$ is
$+1$ for operations that preserve the handedness of the axes and $-1$ for those
that reverse it.

As stated in the main text, the Kerr vector is $\mathcal T$-odd and axial,
whereas strain is a $\mathcal T$-even symmetric polar tensor. Their
transformations at fixed frequency are
\begin{equation}
  \begin{aligned}
    \bm\theta'&=D_\theta(g)\bm\theta,\\
    D_\theta(g)&=(-1)^{t_g}\det(R_g)R_g,\\
    \bm\varepsilon'&=R_g\bm\varepsilon R_g^{\mathrm T}.
  \end{aligned}
  \label{eq:sm_transform}
\end{equation}
Here $D_\theta(g)$ is the transformation (representation) matrix of $\bm\theta$
under $g$. The factor $(-1)^{t_g}$ accounts for the sign of the Kerr vector
under $\mathcal T$, and $\det(R_g)$ accounts for its axial character. Strain is
even under $\mathcal T$, so its transformation depends only on $R_g$.

In the component order specified above, the $6\times6$ matrix $D_\varepsilon(g)$
in Eq. (2) of the main text is obtained by writing the last relation of
Eq.~\eqref{eq:sm_transform} as
\begin{equation}
  \begin{pmatrix}
    \varepsilon'_{xx}&\varepsilon'_{yy}&\varepsilon'_{zz}&\varepsilon
    '_{yz}&\varepsilon'_{zx}&\varepsilon'_{xy}
  \end{pmatrix}^{\mathrm T}
  =D_\varepsilon(g)
  \begin{pmatrix}
    \varepsilon_{xx}&\varepsilon_{yy}&\varepsilon_{zz}&\varepsilon
    _{yz}&\varepsilon_{zx}&\varepsilon_{xy}
  \end{pmatrix}^{\mathrm T}.
\end{equation}
Thus, $D_\varepsilon(g)$ specifies how the six independent strain components mix
or change sign under $g$.

Transforming the Kerr vector obtained from Eq. (1) of the main text gives the
coefficient matrix $D_\theta(g)\Lambda$. Applying the same linear coupling to
the transformed strain instead gives $\Lambda D_\varepsilon(g)$. These two
matrices must agree for arbitrary values of the six independent strain
components, which yields Eq. (2) of the main text \cite{sm001, sm002}. This
constraint determines the vanishing coefficients and the relations among those
that remain. A Kerr component, or a combination of components, can couple
linearly to a strain component or combination belonging to the same irreducible
representation (IRR) of $\mathcal M$.

\subsection{Irreducible decomposition of strain}

The linear combinations in Table~\ref{tab:strain_decomposition} are chosen to
transform within the stated IRRs. The table uses standard crystallographic
point-group settings, including the unique-$b$ setting with the twofold axis
along $y$ for monoclinic groups. Trigonal and hexagonal groups are expressed in
orthonormal Cartesian axes. The strain components must be expressed in the
tabulated axes.

Since strain is $\mathcal T$-even, its representation can be organized by the 32
ordinary crystallographic point groups. For an MPG, the corresponding spatial
group is obtained by omitting the primes that mark operations combined with
$\mathcal T$ and any independent $\mathcal T$ operation $1'$. The $\mathcal T$
factor $(-1)^{t_g}$ must still be retained in $D_\theta(g)$ when Eq. (2) of the
main text is applied.

The table uses real representations of the strain components. A pair of
complex-conjugate one-dimensional representations is combined into a
two-dimensional real representation where applicable. The Mulliken labels $A$
and $B$ denote one-dimensional representations, $E$ two-dimensional
representations, and $T$ three-dimensional representations. The subscript $g$
denotes even parity under spatial inversion. A single or double prime on a
representation label denotes even or odd parity under the relevant mirror
reflection, respectively. These primes describe spatial mirror parity, whereas
primes in MPG symbols mark operations combined with $\mathcal T$. Numerical
subscripts distinguish different IRRs of the same dimension within a point
group. Numerical prefactors in the decomposition give the number of occurrences
of each IRR. Components grouped in parentheses belong to the same
multidimensional IRR. Factors such as 2 in $2\varepsilon_{xy}$ specify the
normalization of the listed linear combinations. Throughout the table,
$\varepsilon_{xy}$ itself denotes tensor shear strain.

\begin{table}
  \caption{Complete decomposition of the six independent components of symmetric
  strain for all 32 crystallographic point groups. The corresponding strain
  components and linear combinations are listed for each irreducible
  representation.}\label{tab:strain_decomposition}
  \footnotesize
  \renewcommand{\arraystretch}{1.15}
  \begin{ruledtabular}
    \begin{tabular}{@{}llp{11.5cm}@{}}
      \shortstack[l]{Crystallographic\\point group} &
      \shortstack[l]{Irreducible decomposition\\of the strain representation} &
      Strain components and
      linear combinations \\
      \colrule
      $1$ & $6A$ & $A$: all six components \\[0.5pt]
      $\bar{1}$ & $6A_g$ & $A_g$: all six components \\[0.5pt]
      $2$ & $4A+2B$ & $A$: $\eps_{xx},\eps_{yy},\eps_{zz},\eps_{zx}$, $B$:
      $\eps_{xy},\eps_{yz}$ \\[0.5pt]
      $m$ & $4A^{\prime}+2A^{\prime\prime}$ & $A^{\prime}$:
      $\eps_{xx},\eps_{yy},\eps_{zz},\eps_{zx}$, $A^{\prime\prime}$:
      $\eps_{xy},\eps_{yz}$ \\[0.5pt]
      $2/m$ & $4A_g+2B_g$ & $A_g$: $\eps_{xx},\eps_{yy},\eps_{zz},\eps_{zx}$,
      $B_g$: $\eps_{xy},\eps_{yz}$ \\[0.5pt]
      $222$ & $3A+B_1+B_2+B_3$ & $A$: $\eps_{xx},\eps_{yy},\eps_{zz}$, $B_1$:
      $\eps_{xy}$, $B_2$: $\eps_{zx}$, $B_3$: $\eps_{yz}$ \\[0.5pt]
      $mm2$ & $3A_1+A_2+B_1+B_2$ & $A_1$: $\eps_{xx},\eps_{yy},\eps_{zz}$,
      $A_2$: $\eps_{xy}$, $B_1$: $\eps_{zx}$, $B_2$: $\eps_{yz}$ \\[0.5pt]
      $mmm$ & $3A_g+B_{1g}+B_{2g}+B_{3g}$ & $A_g$:
      $\eps_{xx},\eps_{yy},\eps_{zz}$, $B_{1g}$: $\eps_{xy}$, $B_{2g}$:
      $\eps_{zx}$, $B_{3g}$: $\eps_{yz}$ \\[0.5pt]
      $4$ & $2A+2B+E$ & $A$: $\eps_{xx}+\eps_{yy},\eps_{zz}$, $B$:
      $\eps_{xx}-\eps_{yy},\eps_{xy}$, $E$: $(\eps_{zx},\eps_{yz})$ \\[0.5pt]
      $\bar{4}$ & $2A+2B+E$ & $A$: $\eps_{xx}+\eps_{yy},\eps_{zz}$, $B$:
      $\eps_{xx}-\eps_{yy},\eps_{xy}$, $E$: $(\eps_{zx},\eps_{yz})$ \\[0.5pt]
      $4/m$ & $2A_g+2B_g+E_g$ & $A_g$: $\eps_{xx}+\eps_{yy},\eps_{zz}$, $B_g$:
      $\eps_{xx}-\eps_{yy},\eps_{xy}$, $E_g$: $(\eps_{zx},\eps_{yz})$ \\[0.5pt]
      $422$ & $2A_1+B_1+B_2+E$ & $A_1$: $\eps_{xx}+\eps_{yy},\eps_{zz}$, $B_1$:
      $\eps_{xx}-\eps_{yy}$, $B_2$: $\eps_{xy}$, $E$: $(\eps_{zx},\eps_{yz})$
      \\[0.5pt]
      $4mm$ & $2A_1+B_1+B_2+E$ & $A_1$: $\eps_{xx}+\eps_{yy},\eps_{zz}$, $B_1$:
      $\eps_{xx}-\eps_{yy}$, $B_2$: $\eps_{xy}$, $E$: $(\eps_{zx},\eps_{yz})$
      \\[0.5pt]
      $\bar{4}2m$ & $2A_1+B_1+B_2+E$ & $A_1$: $\eps_{xx}+\eps_{yy},\eps_{zz}$,
      $B_1$: $\eps_{xx}-\eps_{yy}$, $B_2$: $\eps_{xy}$, $E$:
      $(\eps_{zx},\eps_{yz})$ \\[0.5pt]
      $4/mmm$ & $2A_{1g}+B_{1g}+B_{2g}+E_g$ & $A_{1g}$:
      $\eps_{xx}+\eps_{yy},\eps_{zz}$, $B_{1g}$: $\eps_{xx}-\eps_{yy}$,
      $B_{2g}$: $\eps_{xy}$, $E_g$: $(\eps_{zx},\eps_{yz})$ \\[0.5pt]
      $3$ & $2A+2E$ & $A$: $\eps_{xx}+\eps_{yy},\eps_{zz}$, $E$:
      $(\eps_{xx}-\eps_{yy},2\eps_{xy})$, $(\eps_{zx},\eps_{yz})$ \\[0.5pt]
      $\bar{3}$ & $2A_g+2E_g$ & $A_g$: $\eps_{xx}+\eps_{yy},\eps_{zz}$, $E_g$:
      $(\eps_{xx}-\eps_{yy},2\eps_{xy})$, $(\eps_{zx},\eps_{yz})$ \\[0.5pt]
      $32$ & $2A_1+2E$ & $A_1$: $\eps_{xx}+\eps_{yy},\eps_{zz}$, $E$:
      $(\eps_{xx}-\eps_{yy},2\eps_{xy})$, $(\eps_{zx},\eps_{yz})$ \\[0.5pt]
      $3m$ & $2A_1+2E$ & $A_1$: $\eps_{xx}+\eps_{yy},\eps_{zz}$, $E$:
      $(\eps_{xx}-\eps_{yy},2\eps_{xy})$, $(\eps_{zx},\eps_{yz})$ \\[0.5pt]
      $\bar{3}m$ & $2A_{1g}+2E_g$ & $A_{1g}$: $\eps_{xx}+\eps_{yy},\eps_{zz}$,
      $E_g$: $(\eps_{xx}-\eps_{yy},2\eps_{xy})$, $(\eps_{zx},\eps_{yz})$
      \\[0.5pt]
      $6$ & $2A+E_1+E_2$ & $A$: $\eps_{xx}+\eps_{yy},\eps_{zz}$, $E_1$:
      $(\eps_{zx},\eps_{yz})$, $E_2$: $(\eps_{xx}-\eps_{yy},2\eps_{xy})$
      \\[0.5pt]
      $\bar{6}$ & $2A^{\prime}+E^{\prime}+E^{\prime\prime}$ & $A^{\prime}$:
      $\eps_{xx}+\eps_{yy},\eps_{zz}$, $E^{\prime}$:
      $(\eps_{xx}-\eps_{yy},2\eps_{xy})$, $E^{\prime\prime}$:
      $(\eps_{zx},\eps_{yz})$ \\[0.5pt]
      $6/m$ & $2A_g+E_{1g}+E_{2g}$ & $A_g$: $\eps_{xx}+\eps_{yy},\eps_{zz}$,
      $E_{1g}$: $(\eps_{zx},\eps_{yz})$, $E_{2g}$:
      $(\eps_{xx}-\eps_{yy},2\eps_{xy})$ \\[0.5pt]
      $622$ & $2A_1+E_1+E_2$ & $A_1$: $\eps_{xx}+\eps_{yy},\eps_{zz}$, $E_1$:
      $(\eps_{zx},\eps_{yz})$, $E_2$: $(\eps_{xx}-\eps_{yy},2\eps_{xy})$
      \\[0.5pt]
      $6mm$ & $2A_1+E_1+E_2$ & $A_1$: $\eps_{xx}+\eps_{yy},\eps_{zz}$, $E_1$:
      $(\eps_{zx},\eps_{yz})$, $E_2$: $(\eps_{xx}-\eps_{yy},2\eps_{xy})$
      \\[0.5pt]
      $\bar{6}m2$ & $2A_1^{\prime}+E^{\prime}+E^{\prime\prime}$ &
      $A_1^{\prime}$: $\eps_{xx}+\eps_{yy},\eps_{zz}$, $E^{\prime}$:
      $(\eps_{xx}-\eps_{yy},2\eps_{xy})$, $E^{\prime\prime}$:
      $(\eps_{zx},\eps_{yz})$ \\[0.5pt]
      $6/mmm$ & $2A_{1g}+E_{1g}+E_{2g}$ & $A_{1g}$:
      $\eps_{xx}+\eps_{yy},\eps_{zz}$, $E_{1g}$: $(\eps_{zx},\eps_{yz})$,
      $E_{2g}$: $(\eps_{xx}-\eps_{yy},2\eps_{xy})$ \\[0.5pt]
      $23$ & $A+E+T$ & $A$: $\eps_{xx}+\eps_{yy}+\eps_{zz}$, $E$:
      $(2\eps_{zz}-\eps_{xx}-\eps_{yy},\eps_{xx}-\eps_{yy})$, $T$:
      $(\eps_{yz},\eps_{zx},\eps_{xy})$ \\[0.5pt]
      $m\bar{3}$ & $A_g+E_g+T_g$ & $A_g$: $\eps_{xx}+\eps_{yy}+\eps_{zz}$,
      $E_g$: $(2\eps_{zz}-\eps_{xx}-\eps_{yy},\eps_{xx}-\eps_{yy})$, $T_g$:
      $(\eps_{yz},\eps_{zx},\eps_{xy})$ \\[0.5pt]
      $432$ & $A_1+E+T_2$ & $A_1$: $\eps_{xx}+\eps_{yy}+\eps_{zz}$, $E$:
      $(2\eps_{zz}-\eps_{xx}-\eps_{yy},\eps_{xx}-\eps_{yy})$, $T_2$:
      $(\eps_{yz},\eps_{zx},\eps_{xy})$ \\[0.5pt]
      $\bar{4}3m$ & $A_1+E+T_2$ & $A_1$: $\eps_{xx}+\eps_{yy}+\eps_{zz}$, $E$:
      $(2\eps_{zz}-\eps_{xx}-\eps_{yy},\eps_{xx}-\eps_{yy})$, $T_2$:
      $(\eps_{yz},\eps_{zx},\eps_{xy})$ \\[0.5pt]
      $m\bar{3}m$ & $A_{1g}+E_g+T_{2g}$ & $A_{1g}$:
      $\eps_{xx}+\eps_{yy}+\eps_{zz}$, $E_g$:
      $(2\eps_{zz}-\eps_{xx}-\eps_{yy},\eps_{xx}-\eps_{yy})$, $T_{2g}$:
      $(\eps_{yz},\eps_{zx},\eps_{xy})$ \\[0.5pt]
    \end{tabular}
  \end{ruledtabular}
\end{table}

\clearpage
\section{COMPUTATIONAL METHODS}

First-principles calculations were performed within density functional theory
(DFT) using the Vienna \textit{Ab initio} Simulation Package (VASP)
\cite{sm003}. The electron-ion interactions were described using the projector
augmented-wave (PAW) method \cite{sm004, sm005}, while exchange-correlation
effects were treated using the Perdew-Burke-Ernzerhof (PBE) functional within
the generalized gradient approximation (GGA) \cite{sm006}. The plane-wave energy
cutoff was set to 600 eV after a convergence test, and the Brillouin zone (BZ)
was sampled using \ensuremath{\Gamma}-centered Monkhorst-Pack grids of 9
\ensuremath{\times} 9 \ensuremath{\times} 9, 14 \ensuremath{\times} 14
\ensuremath{\times} 20, and 13 \ensuremath{\times} 13 \ensuremath{\times} 1 for
MnTe\textsubscript{2}, CoF\textsubscript{2}, and the
Cr\textsubscript{2}Se\textsubscript{2}O and
Cr\textsubscript{2}Te\textsubscript{2}O monolayers, respectively \cite{sm007}.
The convergence criteria for total energy and atomic forces were set to
10\textsuperscript{\ensuremath{-}}\textsuperscript{7} eV and 0.001 eV/\AA{},
respectively. The crystal structures of bulk MnTe\textsubscript{2} and
CoF\textsubscript{2} were adopted from experimental reports \cite{sm008, sm009},
whereas those of the Cr\textsubscript{2}Se\textsubscript{2}O and
Cr\textsubscript{2}Te\textsubscript{2}O monolayers were obtained by structural
optimization. The PBE+$U$ method was adopted to capture the localized nature of
the transition-metal 3$d$ electrons \cite{sm010, sm011}, where $U$ denotes the
on-site Coulomb interaction, $J$ the Hund exchange parameter, and
$U_{\mathrm{eff}}=U-J$. For MnTe\textsubscript{2}, $U_{\mathrm{eff}}=4.2$ eV was
used for the Mn 3$d$ orbitals \cite{sm008}. For CoF\textsubscript{2}, $U$ = 4.0
eV and $J$ = 0.6 eV were used for the Co 3$d$ orbitals \cite{sm012}. For the
Cr\textsubscript{2}Se\textsubscript{2}O and
Cr\textsubscript{2}Te\textsubscript{2}O monolayers, the effective interaction
for the Cr 3$d$ orbitals was set to $U_{\mathrm{eff}}$ = 3.5 eV, consistent with
previous studies \cite{sm013, sm014}. For the monolayers, a vacuum space of 15
\AA{} was used to avoid interactions between periodic images. Postprocessing was
performed using VASPKIT to obtain spin-resolved band structures \cite{sm015}.
Magnetic space groups (MSGs) and MPGs were determined using the
magnetic-symmetry programs of the Bilbao Crystallographic Server \cite{sm016},
and the symmetry-adapted optical-conductivity tensors were obtained using
MTENSOR \cite{sm001}. Fully relativistic Bloch states including spin-orbit
coupling (SOC) were transformed into a Wannier representation using Wannier90
\cite{sm017, sm018, sm019}. For MnTe\textsubscript{2}, 200 Bloch bands on a 6
\ensuremath{\times} 6 \ensuremath{\times} 6 mesh were used to construct 112
spinor Wannier functions initialized from Mn 3$d$, 4$s$ and Te 5$s$, 5$p$
orbitals. For CoF\textsubscript{2}, 224 Bloch bands on an 8 \ensuremath{\times}
8 \ensuremath{\times} 12 mesh were used to construct 56 spinor Wannier functions
initialized from Co 3$d$, 4$s$ and F 2$s$, 2$p$ orbitals. For the
Cr\textsubscript{2}Se\textsubscript{2}O and
Cr\textsubscript{2}Te\textsubscript{2}O monolayers, the Wannierization employed
182 Bloch bands on a 25 \ensuremath{\times} 25 \ensuremath{\times} 1 $k$-point
mesh and 38 spinor Wannier functions initialized from the Cr 3$d$, Se 4$p$ (or
Te 5$p$), and O 2$p$ orbitals.

The frequency-dependent complex optical-conductivity tensor was evaluated from
the Wannier-interpolated Hamiltonian using the Kubo-Greenwood implementation in
postw90 \cite{sm019, sm020, sm021, sm022}. The interband contribution is given
by

\begin{equation}
  \begin{aligned}
    \sigma_{\alpha\beta}(\omega,\bm\varepsilon)
    ={}&\frac{ie^2\hbar}{N_k\varOmega_{\mathrm c}}\sum_{\bm k}\sum_{n\ne m}
    \frac{f_{m\bm k}-f_{n\bm k}}{\varepsilon_{m\bm k}-\varepsilon_{n\bm k}}\\
    &\times\frac{\langle\psi_{n\bm k}|v_\alpha|\psi_{m\bm k}\rangle
    \langle\psi_{m\bm k}|v_\beta|\psi_{n\bm k}\rangle}
    {\varepsilon_{m\bm k}-\varepsilon_{n\bm k}-(\hbar\omega+i\eta)}.
  \end{aligned}
\end{equation}

Here, $i$ is the imaginary unit, $\omega$ is the angular frequency and $\bm k$
is the Bloch wave vector. The indices $\alpha$ and $\beta$ denote Cartesian
directions, $e$ is the elementary charge, $\hbar$ is the reduced Planck
constant, and $N_{k}$ is the number of sampled $k$ points. The indices $n$ and
$m$ label Bloch bands. $f_{n\bm k}$, $\varepsilon_{n\bm k}$, and $|\psi_{n\bm
k}\rangle$ are their occupation factors, energies, and eigenstates,
respectively. $v_{\alpha }$ is the $\alpha$ component of the velocity operator.
For bulk MnTe\textsubscript{2} and CoF\textsubscript{2}, $\varOmega_{\mathrm c}$
is the unit-cell volume. For Cr\textsubscript{2}Se\textsubscript{2}O and
Cr\textsubscript{2}Te\textsubscript{2}O monolayers, it is the effective layer
volume, given by the product of the in-plane unit cell area and the layer
thickness. The parameter $\eta$ denotes the energy broadening.

The broadening parameter was fixed at $\eta=0.1$ eV. The BZ integration was
performed on 61 \ensuremath{\times} 61 \ensuremath{\times} 61, 61
\ensuremath{\times} 61 \ensuremath{\times} 91, and 521 \ensuremath{\times} 521
\ensuremath{\times} 1 meshes for MnTe\textsubscript{2}, CoF\textsubscript{2},
and the monolayers, respectively, with the photon energy sampled at an interval
of 0.01 eV. Throughout the optical calculations, the propagation direction and
incident polarization were fixed at $\bm q\parallel z$ and $\bm E\parallel x$,
respectively, where $\bm q$ and $\bm E$ denote the incident-light wave vector
and electric field.

The antisymmetric optical conductivity introduced in the main text is
\begin{equation}
  \sigma^{\mathrm A}_{xy}(\omega,\bm\varepsilon)
  =\frac{\sigma_{xy}(\omega,\bm\varepsilon)-\sigma_{yx}(\omega,\bm\varepsilon
  )}{2}.
\end{equation}

\clearpage
\section{PIEZO-KERR EFFECT IN MONOLAYER
Cr\textsubscript{2}Te\textsubscript{2}O}

\begin{figure}[H]
  \centering
  \includegraphics[width=5.25in]{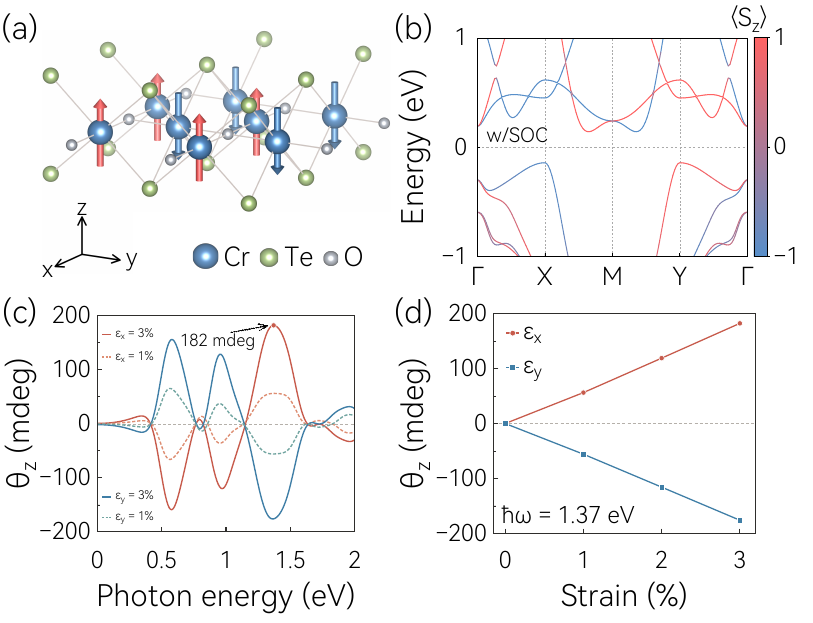}
  \caption{(a) Crystal structure and magnetic configuration of monolayer
  Cr\textsubscript{2}Te\textsubscript{2}O. Arrows denote magnetic moments. (b)
  Band structure with SOC and without strain. (c) Kerr rotation $\theta_z$ under
  uniaxial tension along $x$ and $y$ directions. (d) Strain dependence of
  $\theta_z$ at a photon energy of $\hbar\omega=1.37$ eV.}
  \label{fig:S1}
\end{figure}

Cr\textsubscript{2}Te\textsubscript{2}O provides a second monolayer
verification. The unstrained band structure calculated with SOC ($\bm N\parallel
z$) shows that Cr\textsubscript{2}Te\textsubscript{2}O is semiconducting and
exhibits momentum-dependent spin splitting [Fig.~\ref{fig:S1}(b)]. Here, $\bm N$
is the N\'eel vector. The Kerr angle $\theta_z$ vanishes in the unstrained
monolayer but becomes nonzero under uniaxial tension. It reverses sign when
equal uniaxial tensile strains are switched between the $x$ and $y$ axes
[Fig.~\ref{fig:S1}(c)]. Uniaxial tensile strain can also enhance the magnitude
of the Kerr angle [Fig.~\ref{fig:S1}(d)]. At a fixed photon energy of
$\hbar\omega=1.37$ eV, $|\theta_z|$ reaches approximately 182 mdeg at 3\%
tension [Fig.~\ref{fig:S1}(d)].

\clearpage
\section{ELECTRONIC STRUCTURES}

\begin{figure}[H]
  \includegraphics[width=\linewidth]{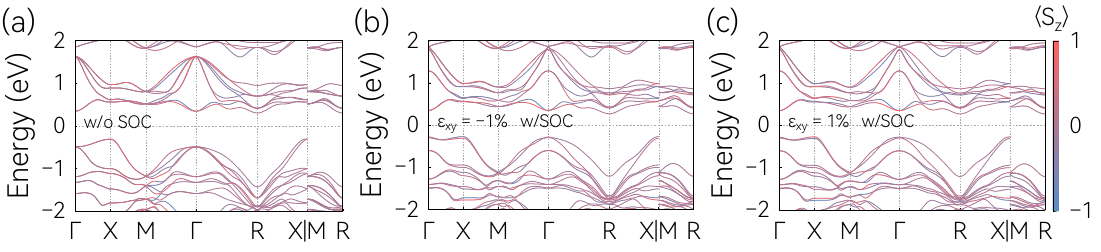}
  \caption{Band structures of MnTe\textsubscript{2} (a) without strain or SOC
  and (b), (c) with SOC under $\varepsilon_{xy}=-1\%$ and $1\%$, respectively.
  The noncoplanar magnetic configuration follows Fig.~2(a) of the main text.
  Colors represent the normalized spin projection along the $z$ axis.}
  \label{fig:S2}
\end{figure}

For MnTe\textsubscript{2}, spin splitting is already present without SOC
[Fig.~\ref{fig:S2}(a)], consistent with its reported nonrelativistic spin
splitting in a noncoplanar antiferromagnetic (AFM) configuration \cite{sm008}.
With SOC for this noncoplanar magnetic configuration, MnTe\textsubscript{2}
remains semiconducting under $\varepsilon_{xy}=\pm1\%$ [Figs.~\ref{fig:S2}(b)
and \ref{fig:S2}(c)]. The two opposite shear strains yield band curves with
similar overall shapes.

\begin{figure}[H]
  \includegraphics[width=\linewidth]{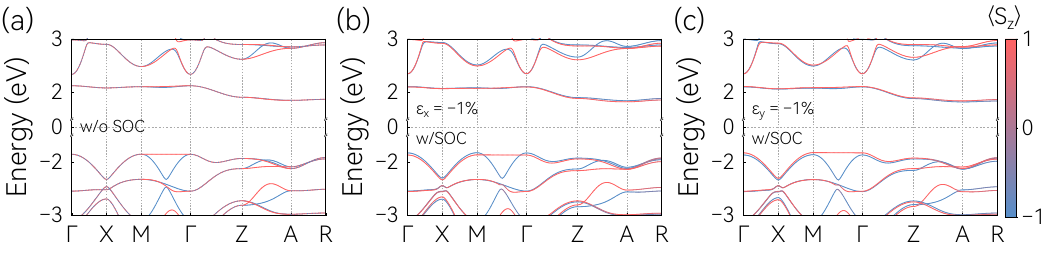}
  \caption{Band structures of CoF\textsubscript{2} (a) without strain or SOC and
  (b), (c) with SOC under 1\% uniaxial compression along $x$ and $y$,
  respectively. Colors represent the normalized spin projection along the $z$
  axis.}
  \label{fig:S3}
\end{figure}

The unstrained CoF\textsubscript{2} bands also exhibit spin splitting without
SOC [Fig.~\ref{fig:S3}(a)]. Under 1\% uniaxial compressive strains along the $x$
and $y$ axes, CoF\textsubscript{2} remains insulating with SOC ($\bm N\parallel
z$) [Figs.~\ref{fig:S3}(b) and \ref{fig:S3}(c)].

\begin{figure}[H]
  \includegraphics[width=\linewidth]{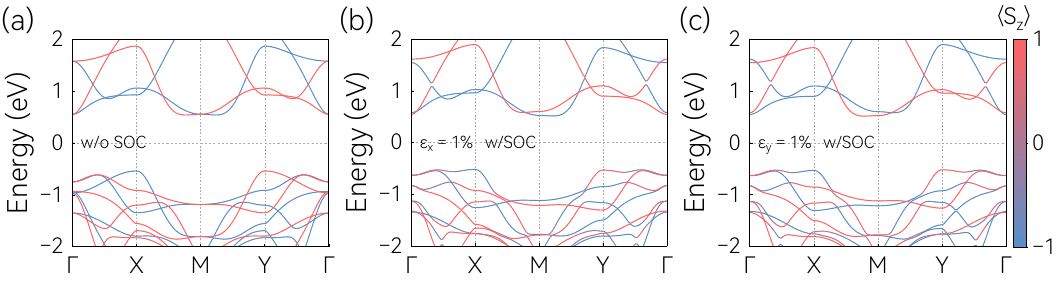}
  \caption{Band structures of monolayer Cr\textsubscript{2}Se\textsubscript{2}O
  (a) without strain or SOC and (b), (c) with SOC under 1\% uniaxial tension
  along $x$ and $y$, respectively. Colors represent the normalized spin
  projection along the $z$ axis.}
  \label{fig:S4}
\end{figure}

\begin{figure}[H]
  \includegraphics[width=\linewidth]{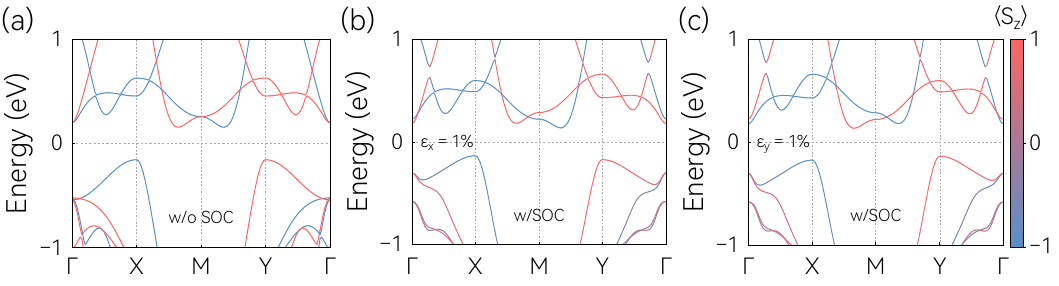}
  \caption{Band structures of monolayer Cr\textsubscript{2}Te\textsubscript{2}O
  (a) without strain or SOC and (b), (c) with SOC under 1\% uniaxial tension
  along $x$ and $y$, respectively. Colors represent the normalized spin
  projection along the $z$ axis.}
  \label{fig:S5}
\end{figure}

The spin-resolved band structures without SOC show that monolayer
Cr\textsubscript{2}Se\textsubscript{2}O and
Cr\textsubscript{2}Te\textsubscript{2}O are indirect-gap semiconductors. Despite
their zero net magnetic moments, Cr\textsubscript{2}Se\textsubscript{2}O and
Cr\textsubscript{2}Te\textsubscript{2}O exhibit pronounced momentum-dependent
spin splitting without SOC, consistent with their altermagnetic character.
Figures~\ref{fig:S4}(a) and \ref{fig:S5}(a) show the unstrained bands of
Cr\textsubscript{2}Se\textsubscript{2}O and
Cr\textsubscript{2}Te\textsubscript{2}O, respectively. Figures~\ref{fig:S4}(b),
\ref{fig:S4}(c), \ref{fig:S5}(b), and \ref{fig:S5}(c) present the bands with SOC
($\bm N\parallel z$) under 1\% uniaxial tensile strains along the $x$ and $y$
axes. Both monolayers remain semiconducting under the strains shown. The labels
$\varepsilon_x$ and $\varepsilon_y$ denote uniaxial strains applied along the
$x$ and $y$ directions, respectively.

\clearpage
\bibliography{SM/Supplemental_Material-reference}